\documentclass[aps,preprint]{revtex4}
\usepackage{graphicx}
\usepackage{rotating}
\usepackage{dcolumn}
\usepackage{epsfig}

\usepackage{latexsym}
\usepackage{mathrsfs}
\usepackage{amssymb}
\usepackage{amsmath}
\usepackage{amsfonts}

\newtheorem{teorema}{Theorem}
\newtheorem{propo}{Proposition}
\newtheorem{lemma}{Lemma}

\newtheorem{defi}{Definition}

\begin{document}

\title{General Framework for phase synchronization through localized sets}
\author{T.   Pereira,  M.S. Baptista, and J. Kurths}

\affiliation{
Nonlinear Dynamics, Institute of Physics,
University of Potsdam, D-14415, Potsdam, Germany
}

\date{\today}

\begin{abstract}
  
  We present an approach which enables to identify phase
  synchronization in coupled chaotic oscillators without having to
  explicitly measure the phase.  We show that if one defines a typical
  event in one oscillator and then observes another one whenever this
  event occurs, these observations give rise to a localized set.  Our
  result provides a general and easy way to identify PS, which can
  also be used to oscillators that possess multiple time scales.  We
  illustrate our approach in networks of chemically coupled neurons.
  We show that clusters of phase synchronous neurons may emerge before
  the onset of phase synchronization in the whole network, producing a
  suitable environment for information exchanging.  Furthermore, we
  show the relation between the localized sets and the amount of
  information that coupled chaotic oscillator can exchange.

\end{abstract}

\maketitle

\section{Introduction}

The emergency of collective behavior among coupled oscillators is a
rather common phenomenon.  In nature, one typically finds interacting
chaotic oscillators which through the coupling scheme form small and
large networks. Surprisingly, even though chaotic systems possess an
exponential divergency of nearby trajectories, they can synchronize
due to the coupling, still preserving the chaotic
behavior\cite{fujisaka,pecora90,pecora98}.  Indeed, synchronization
phenomena have been found in a variety of fields as ecology
\cite{blasius}, neuroscience \cite{reynaldo,juergen,thiel}, economy
\cite{economy}, and lasers \cite{imaculada,laser,roy}.

In the last years some types of synchronization have been reported
\cite{livro}.  A rather interesting kind is a weak synchronization,
namely phase synchronization (PS), that does not reveal itself
directly from the trajectory, but as a boundedness of phase difference
between the interacting oscillators. In such a synchronization the
trajectories can be uncorrelated, and therefore, the oscillators
present some independence of the amplitudes, but still preserving the
collective behavior.

This phenomenon can arise from a very small coupling strength
\cite{rosenblum}. It has been reported that it mediates processes of
information transmission and collective behavior in neural and active
networks \cite{Murilo-Canal}, and communication processes in the human
brain \cite{fell:2002,mormann:2003}. Its presence has been found in a
variety of experimental systems, such as in electronic circuits
\cite{parlitz,baptista:2003}, in electrochemical oscillators
\cite{hudson}, plasma physics \cite{Epa}, and climatology
\cite{douglas}.

In order to state the existence of PS, one has to introduce a phase
$\phi(t)$ for the chaotic oscillator, what is not straightforward.
Even though the phase is expected to exist to a general attractor, due
to the existence of the zero Lyapunov exponent \cite{livro}, its
explicit calculation may be impossible. Actually, even for the simple
case of coherent attractors, it has been shown that phases can be
defined in different ways, each one being chosen according to the
particular case studied.  However, all of them agree for sufficiently
coherent attractors \cite{Josic}.

In spite of the large interest in this field, there is still no
general, systematic, and easy way to detect the existence of this
phenomenon, mainly, due to the fact that the phase is rather difficult
(often unknown) to calculate. The calculation becomes even harder if
the oscillators are non-coherent, e.g. the funnel oscillator
\cite{livro}.  Therefore, in order to present a general
approach to detect PS, with practical applications, we must overcome
the need of a phase.

In many cases the phase can be estimated via the Hilbert
transformation or a wavelet decomposition \cite{livro}.  Supposing
that it is possible to get a phase, the approach developed in Ref.
\cite{nature} gives rather good results.  It is grounded on the idea
of conditional observations of the oscillators. Whenever the phase of
one of the oscillators is increased by $2\pi$, we measure the phase of
the other oscillators.  The main idea is that if one has PS, the
distribution of these conditional observation in the phase presents a
sharp peak, and therefore PS can be detected.
  
There are a few approaches that try to overcome the difficulties of
not having a general phase. For periodically driven oscillators, there
is an interesting approach, very useful and easy to implement that
overcomes the need of a phase, the stroboscopic map technique.  It
consists in sampling the chaotic trajectory at times $n T_0$, where
$n$ is an integer and $T_0$ is the period of the driver.  The
stroboscopic map was used to detect PS \cite{livro,Epa,baptista:2003}.
The basic idea is that if the stroboscopic map is localized in the
attractor, PS is present.  Actually, the stroboscopic map is a
particular case of the approach of Ref.  \cite{nature}. Indeed, since
the driver is periodic, the observation of the trajectory of the
chaotic oscillators at times $n T_0$ is equivalent to observe the
oscillators at every increasing of $2\pi$ in the phase of the driver.
Furthermore, if the chaotic oscillator presents a sharp conditional
distribution, this means that the stroboscopic map is localized. The
advantage of such an approach is that it does not require the
introduction of a phase neither in the periodic oscillator nor in the
chaotic one.

In the case of two or more coupled chaotic oscillators, namely
$\Sigma_j$ and $\Sigma_k$, the stroboscopic map techniques can be no
longer applied. However, if the oscillators are coherent and have a
proper rotation, a generalization of the stroboscopic map has been
recently developed \cite{PHD}. Instead of observing the oscillators at
fixed time intervals, multiples of the period, one can define a
Poincar\'e section in $\Sigma_j$ and then observe $\Sigma_k$ every
time the trajectory of $\Sigma_j$ crosses the Poincar\'e section. If
the oscillators are in PS, these observations give place to a
localized set.

Another approach that is relevant to the present problem is the one
developed in Ref. \cite{quio}.  This approach consists of defining a
point ${\bf x}_j(t) \in \Sigma_j$ and a small neighborhood of this
point composed by points ${\bf x}_j(t_i) \in \Sigma_j$, where
$i=1,\ldots,N$, with $N$ being the number of points within the defined
neighborhood. Then, one observes the oscillator $\Sigma_k$ at the
times $t_i$, which gives place to the points ${\bf
  x}_k(t_i)\in\Sigma_k$.  Again, the idea is that if the oscillators
present synchronization, the cloud of points ${\bf x}_k(t_i)$ occupies
an area much smaller than the attractor area. Further, estimators have
been introduced to quantify the amount of synchronization \cite{quio}.

Even though the intuition says that localized set implies the presence
of synchronization, there is a lack of theoretical analysis showing
such a result for a general oscillator. Moreover, as far as we know,
there are no results that guarantee that such an approach works for
multiple time-scale oscillators. In addition, it is not clear what
kind of points (events) could be chosen, and finally, how one should
proceed in the case that the small neighborhood of the point ${\bf
  x}_j(t) \in \Sigma_j$ has infinitely many neighbor points.

In this work, we extent the ideas of Ref.
\cite{livro,Epa,baptista:2003,PHD,quio}. We show that all these
approaches can be put in the framework of localized sets.  Our results
demonstrate that for general coupled oscillators $\Sigma_j$ and
$\Sigma_k$, if one defines a typical event in $\Sigma_j$ and then
observes the oscillator $\Sigma_k$ whenever this event occurs, these
observations give rise to a localized set in the accessible phase
space if PS exists.  These results can be applied to oscillators that
possess multiple time-scales as well as in neural networks. As an
application, we analyze the onset of PS in neural networks. We show
that in general neural networks one should expect to find clusters of
phase synchronized neurons that can be used to transmit information in
a multiplexing and multichannel way.  Finally, we relate the localized
sets from our theory to the information exchange between the coupled
chaotic oscillators.

The paper is organized as follows: In Sec. \ref{Setup} we define the
dynamical systems we are working on. In Sec. \ref{localized} we give a
result that enables the identification of PS without having to measure
the phase. We illustrate these findings with two coupled R\"ossler
oscillators in Sec.  \ref{ross}. For oscillators possessing multiple
time-scales our main results are discussed in Sec. \ref{timescales},
and then illustrated in Sec. \ref{neus} for bursting neurons coupled
via inhibitory synapses. Our results are also applied to neural
networks of excitatory neurons in Sec.  \ref{net}. We briefly discuss
how to apply these ideas into high dimension oscillators and
experimental data series in Sec. \ref{High}.  Finally, we analyze the
relation between the localized sets and the transmission of
information in chaotic oscillators in Sec. \ref{info}. Moreover, in
Appendix \ref{proof} we prove the main theorem of Sec.
\ref{localized} about the localization of sets in PS.

\section{Basic Set up}\label{Setup}

We consider $N$ oscillators given by first order coupled differential
equations:
\noindent
\begin{equation}
\dot{{\bf x}}_i = {\bf F}_i({\bf x}_i) + \sum_{j=1}^{N}C_{ij} {\bf H}_j({\bf x}_j,{\bf x}_i) 
\label{ds}
\end{equation}
\noindent
where, ${\bf x}_i \in \mathbb{R}^n_i$, and ${\bf F}_i:\mathbb{R}^{n_i}
\rightarrow \mathbb{R}^{n_i}$, ${\bf H}_j$ is the output vector
function, and $C_{ij}$ is the coupling strength between $j$ and $i$.
Note that $C_{ij}$ could also depend on the coordinates and on time.
From now on, we shall label the coupled oscillator ${\bf x}_i$ by
subsystem $\Sigma_i$. Next, we assume that each $\Sigma_j$ has a
stable attractor, i.e. an inflowing region of the phase space where
the solution of $\Sigma_j$ lies. Further, we assume that the subsystem
$\Sigma_j$ admits a phase $\phi_j(t)$. Therefore, the condition for PS
between the oscillators $\Sigma_j$ and $\Sigma_k$ can be written as:
\begin{equation}
|m \phi_j(t) - n \phi_k(t)| < c,                     
\end{equation}
where $n$ and $m$ are integers, and the inequality must hold for all
times, with $c$ being a finite number. For a sake of simplicity, we
consider the case where $n=m=1$, in other words $1:1$ PS. Herein, we
suppose that a frequency $\Omega_j$ can be defined in each subsystem
$\Sigma_j$, such that:
\begin{equation}
\dot{\phi}_j = \Omega_j({\bf x}_1,\cdots,{\bf x}_N,t),
\end{equation}  
where $\Omega_j$ is a continuous function bounded away from zero.
Furthermore there is a number $M$ such that $\Omega_j \le M$.  This
phase is an abstract phase in the sense that it is well defined, but
we are not able to write the function $\Omega_j$ for a general
oscillator. We also consider the frequencies $\dot{\phi}_j$ not to be
too different, such that, in general, through the coupling PS can be
achieved.

\section{Localized Sets in PS states}\label{localized}

In this section we present our main result. The basic idea consists in
the following: Given two subsystems $\Sigma_k$ and $\Sigma_j$, we
observe $\Sigma_k$ whenever an event in the oscillator $\Sigma_j$
happens. As a consequence of these conditional observations, we get a
set $\mathcal{D}_k$. Depending on the properties of this set one can
state whether there is PS.

The conditional observations could be given by a Poincar\'e section,
if it is possible to define a Poincar\'e Section with the property
that the trajectory crosses it once per cycle in a given direction.
We wish to point out that in this case, one is able to have more
information about the dynamics and the phase synchronization
phenomenon. As an example, one can introduce a phase, and estimate the
average frequency of the oscillators. However, these techniques based
on the Poincar\'e section \cite{livro,PHD} cannot be applied to
attractors without a proper rotation, where such a section cannot be
well defined.

\begin{figure}[!h]
\centerline{\hbox{\psfig{file=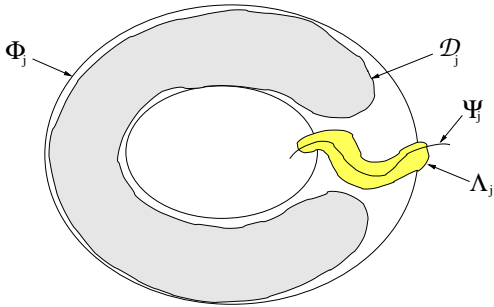,height=4cm}}}
\caption{(Color online) An illustration of the Def. \ref{loc}. The set $\mathcal{D}_j$ does not 
  intersect the neighborhood $\Lambda_j$, therefore, $\mathcal{D}_j$ is a
  localized set of $\Phi_j$.}
\label{defi}
\end{figure}

Our main result overcomes the need of a Poincar\'e Section. We show
that one can use any typical event to detect PS. Such events may be
the crossing of the trajectory with a small piece of a Poincar\'e
section (when it is possible to defined such a section), the crossing
of the trajectory with an arbitrary small segment, the entrance of the
trajectory in an $\varepsilon$-ball, and so on.  The only
constraint is that the event must be typical (we shall clarify what
we mean by typical, later on ) and the region where the event is
defined must have a positive measure.  Let $(t_{k,j}^i)_{i\in\mathbb{N}}$ 
be the time at which the $i$th event in the subsystem $\Sigma_{k,j}$ 
happens.  Then, we construct the set:
\begin{equation}
\mathcal{D}_k \equiv \cup_{i\in\mathbb{N}}{\bf F}^{t_j^i}({\bf x}_k^0),
\end{equation}
where ${\bf x}_k^0$ is the initial point within the attractor of
$\Sigma_k$.  Next, we define what we understand by localized set.
\begin{defi}
  Let $\mathcal{D}_j$ be a subset of $\Phi_j$. The set $\mathcal{D}_j$
  is localized in $\Phi_j$ if there is a cross section $\Psi_j$ and a
  neighborhood $\Lambda_j$ of $\Psi_j$ such that $\mathcal{D}_j \cup
  \Lambda_j = \emptyset$
\label{loc}
\end{defi}
An illustration of the definition is given in Fig. \ref{defi}.

Under the assumptions of Sec. \ref{Setup}, the following result
connects the existence of phase synchronization with the localization
of sets the $\mathcal{D}$:
\begin{teorema}
  Given a typical event, with positive measure, in the oscillator
  $\Sigma_j$, generating the times $(t_j^i)_{i\in \mathbb{N}}$. The
  observation of $\Sigma_k$ at $(t_j^i)_{i\in \mathbb{N}}$ generates a
  localized set $\mathcal{D}_k$ if there is PS.
\end{teorema}

This result constitutes a direct generalization of approaches of Refs.
\cite{quio,PHD}.  As a consequence, this result shed a light into the
problem of PS detection, which turned out to be a rather difficult
task, depending on the system faced. Therefore, PS can be detected in
real-time experiments and in data analysis by verifying whether the
sets $\mathcal{D}$ are localized, without needing any further
calculations.

\subsection{Connection between $\mathcal{D}$ and Unstable Periodic Orbits}

In this section we investigate the mechanism for the non localization
of the sets $\mathcal{D}$.  We let the event definition be an entrance
in an $\varepsilon$-ball in both subsystems, with $\varepsilon$ being
the radius. When $\varepsilon$ is small enough, we can demonstrate
that PS leads to the locking of all unstable periodic orbits (UPO)
between the subsystems.

\begin{propo}
  If the set $\mathcal{D}_k$ is localized, then all UPOs between
  $\Sigma_k$ and $\Sigma_j$ are locked.
\end{propo}

\begin{figure}[h]
  \centerline{\hbox{\psfig{file=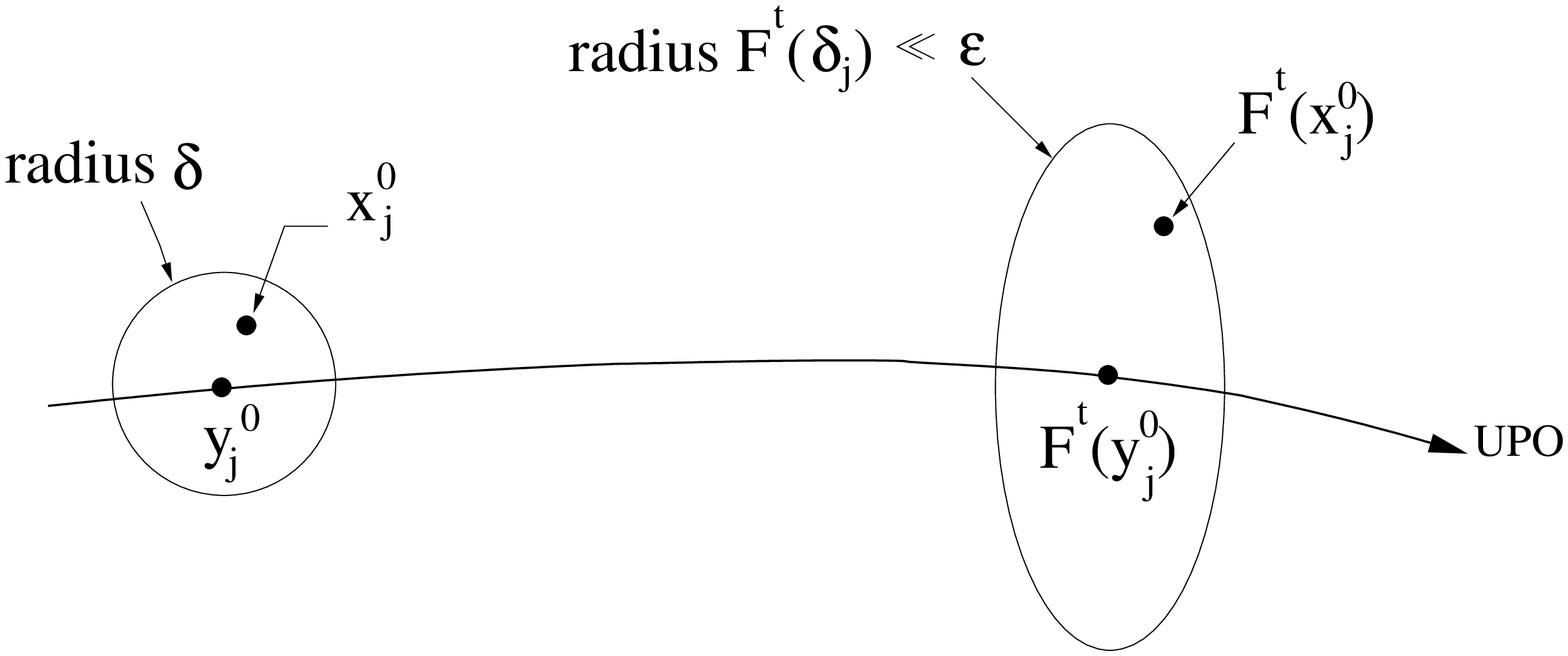,width=8.0cm}}}
\caption{ Illustration of the dynamics near a UPO.}
\label{UPO}
\end{figure}

{\bf Proof:} We demonstrate this result by absurd. Let us assume that
there is PS; as a consequence the set $\mathcal{D}_k$ is localized.
Suppose that there is an UPO, regarded as $\mathcal{X}_j$ in
$\Sigma_j$, and another UPO, regarded as $\mathcal{X}_k$ in
$\Sigma_k$, and that they are not locked (there is no rational number
that relates both frequencies).  So, there is a mismatch between the
frequencies of the two UPOs.  Given an $\varepsilon_j$-ball around
${\bf y}_{j}^0$ (resp.  ${\bf y}_{k}^0$), where ${\bf y}_{j}^0 \in
\mathcal{X}_j$ (resp.  ${\bf y}_{k}^0 \in \mathcal{X}_k$), any point
${\bf x}_j$ distant $\delta_j$ from ${\bf y}_{j}^0$, where $\delta_j
\ll \varepsilon_j$, follows $d [{\bf F}_j^t({\bf x}_j),{\bf
  F}_j^t({\bf y}_j^0)] \le \varepsilon_j$, for any $t \le \tilde t
\approx \ell n (\varepsilon_j / \delta_j) / \lambda_{max}$, where
$\lambda_{max}$ is the largest eigenvalue associated with the orbit
$\mathcal{X}_j$, and $d[ \cdot, \cdot]$ is a metric.  An initial
condition inside the $\delta_j$-ball is governed by the UPO
$\mathcal{X}_j$ till a time $t \le \tilde t$, see Fig.  \ref{UPO} for
an illustration.
Next, we construct the set $\mathcal{D}_k$ by sampling the trajectory
of $\Sigma_k$ whenever the trajectory $\Sigma_j$ enters in the
$\varepsilon$-ball, which is equivalent to observe $\Sigma_k$ every
period of the UPO $\mathcal{X}_j$. There is an one-to-one
correspondence (isomorphism) between the dynamics of the conditional
observations and the dynamics of the irrational rotation in the
unitary circle, $R_{\alpha}: S^1 \rightarrow S^1$, $R_{\alpha} =
e^{\alpha \sqrt{-1}}z$, where $\alpha$ is the frequency mismatch
between the two UPOs, here given by:
\noindent
\begin{equation}
\alpha = inf_{a,b}\{ a \times \omega_{\mathcal{X}_j}  - b \times \omega_{\mathcal{X}_k} \},
\end{equation}
\noindent
where $\omega_{\mathcal{X}_{j,k}}$ is the angular frequency of
$\mathcal{X}_{j,k}$. This means that the points of $\mathcal{D}_k$
will be dense around the UPO $\mathcal{X}_k$, and therefore, the set
$\mathcal{D}_k$ is not localized; there is no PS, what contradict our
assumption.  Indeed, since $\Delta \omega \ge 0$, it is impossible to
bound the phase difference between $\Sigma_j$ and $\Sigma_k$ by a
finite number. Thus, in order to have localized sets $\mathcal{D}$,
all UPOs must be locked.$\Box$

This shows that the mechanism for the non-localization of the sets
$\mathcal{D}$ will be the existence of unlocked UPOs between
$\Sigma_j$ and $\Sigma_k$. Similar results have been pursued for
periodically driven oscillators, \cite{livro}. Right at the
desynchronization some UPOs become unlocked and the stroboscopic map
becomes non-localized, and some phase slips happen, generating an
intermittent behavior. The duration of the phase slips are related to
the number of unlocked UPOs. Of course, in this regime the set
$\mathcal{D}$ is a non-localized set. However, if one looks for finite
time intervals the set $\mathcal{D}$ may be apparently localized.

\section{Coupled R\"ossler Oscillators}\label{ross}

We first illustrate this result for two coupled R\"ossler oscillators,
given by:
\begin{eqnarray}
\dot{x}_{1,2} &=& -\alpha_{1,2}y_{1,2}-z_{1,2}+\epsilon
(x_{2,1}-x_{1,2}), \nonumber \\
\dot{y}_{1,2} &=& \alpha_{1,2}x_{1,2}+0.15y_{1,2}, \label{Rossler}\\
\dot{z}_{1,2} &=& 0.2+z_{1,2}(x_{1,2}-10), \nonumber 
\end{eqnarray}
with $\alpha_{1}=1$, and $\alpha_{2}=\alpha_{1}+\delta \alpha_2$.  In
such a coherent oscillator, we can simply define a phase $tan \phi_i =
y_i/x_i$, where $i=1,2$, which provides an explicity equation for it.
Indeed, taking the derivative with respect to time:
\begin{equation}
\frac{\partial}{\partial \phi_i} tan(\phi_i) \times \dot{\phi}_i = \frac{d}{dt}\frac{y_i}{x_i},
\label{dphase}
\end{equation}
\noindent
which can be written as $ sec^2(\phi_i) \times \dot{\phi}_i =
(\dot{y}_i x_i - y_i \dot{x}_i)/x_i^2$, which provides:
\begin{equation}
\phi_i(t) = \int_0^t \frac{\dot{y}_i x_i - \dot{x}_i y_i}{x^2_i + y^2_i} dt,
\label{phase}
\end{equation}
\noindent
noting that $sec^2\phi_i = (x^2_i + y^2_i)/x^2_i$. In a more compact
notation, we consider ${\bf x}_i = (x_i,y_i)$, then Eq. (\ref{phase})
can be written as
\begin{equation}
\phi_i(t) = \int_0^t \frac{ \dot{\bf x}_i \wedge {\bf x}_i }{|{\bf x}_i|^2} dt,
\label{ph}
\end{equation}
where $\wedge$ represents the vectorial product. Equation (\ref{ph})
can be used to calculate the phase of the oscillators $\Sigma_i$, and
there is PS if $\Delta \phi = \phi_2 - \phi_1$ remains bounded as $t
\rightarrow \infty$.

In order to apply our results we may define an event occurrence in
both oscillators. We define the event in oscillator $\Sigma_1$ to be
the trajectory crossing with the segment:
\begin{equation}
\mathcal{S}_1 = \{ x_1,y_1,z_1 \in \mathbb{R} | x_1 < -13, y_1=0, \mbox{ and } \dot{y_1}>0 \}, 
\end{equation}
the crossings generate the times $(t_1^i)_{i\in \mathbb{N}}$.  The
event in the oscillator $\Sigma_2$ happens whenever its trajectory
crosses the segment:
\begin{equation} 
\mathcal{S}_2 = \{ x_2,y_2,z_2 \in \mathbb{R} |  x_2 > 5 , y_2=10,  \mbox{ and } \dot{y_2}<0\}, 
\end{equation}
the crossings generates the times $(t_2^i)_{i\in \mathbb{N}}$. Then,
the set $\mathcal{D}_{2,1}$ is constructed by observing the
oscillators $\Sigma_{2,1}$ at times $(t_{1,2}^i)_{i\in\mathbb{N}}$
\begin{figure}[!h]
  \centerline{\hbox{\psfig{file=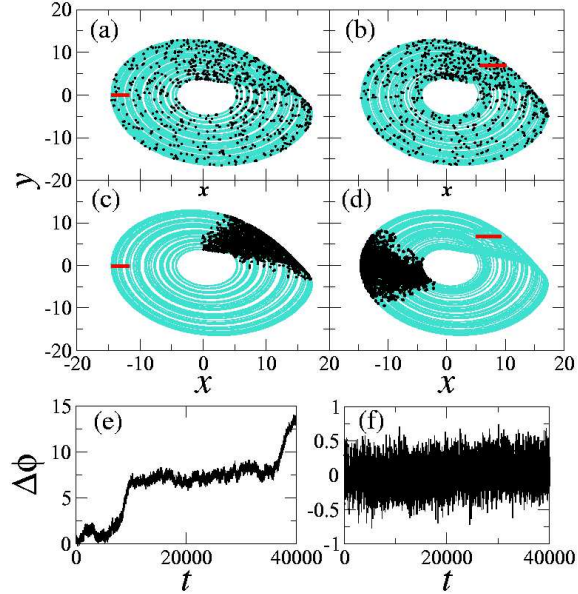,height=8cm}}}
\caption{(Color online) PS onset in two coupled R\"ossler oscillators.
  In (a,c) we depict the attractor of the oscillator $\Sigma_1$ and in
  (b,d) the attractor of $\Sigma_2$ in light gray, the sets
  $\mathcal{D}$ are depicted in black. The bars on Figs. (a) and (c)
  represent the segment $\mathcal{S}_1$, while in Figs. (b) and (d)
  the segment $\mathcal{S}_2$.  In (a) and (b) the sets
  $\mathcal{D}_1$ and $\mathcal{D}_2$ spread over the attractor of the
  oscillator $\Sigma_1$ and $\Sigma_2$, respectively; and there is no
  PS, the phase difference diverges (e).  The parameters are $\epsilon
  = 0.001$ and $\Delta \alpha = 0.001$.  In (c) and (d) the sets
  $\mathcal{D}_1$ and $\mathcal{D}_2$, respectively, are localized and
  there is PS; the phase difference is bounded (f). The parameters are
  $\epsilon = 0.011$ and $\Delta \alpha = 0.001$.}
\label{event}
\end{figure}
\noindent

For $\epsilon = 0.001$ and $\Delta \alpha_2 = 0.001$, the set
$\mathcal{D}_1$ spreads over the attractor of $\Sigma_1$ [ Fig
\ref{event} (a)], and $\mathcal{D}_2$ spreads over the attractor of
$\Sigma_2$ [ Fig.  \ref{event}(b)]. Therefore, there is no PS, i.e.
the phase difference $\Delta \phi$ diverges [Fig \ref{event} (e)].
Indeed, a calculation of the frequencies shows that $\langle
\dot{\phi}_1 \rangle =1.03479$ and $\langle \dot{\phi}_2 \rangle
=1.03508$.  As we increase the coupling, PS appears. In particular,
for $\epsilon = 0.011$ and $\Delta \alpha = 0.001$, the sets
$\mathcal{D}_1$ and $\mathcal{D}_2$ are localized [Figs \ref{event}
(c) and (d), respectively]. Hence, the phase difference is bounded
[Fig.  \ref{event}(f)]. The average frequency is $\langle \dot{\phi}_1
\rangle = \langle \dot{\phi}_2 \rangle = 1.03522$.

\subsection{Estimating the synchronization level}

Our main goal is to state the existence of PS, however, we can also
estimate the synchronization level between $\Sigma_j$ and $\Sigma_k$
by means of the localized sets.  This can be done by introducing an
estimator $H_{jk}$. One way to estimate the amount of synchrony is to
define:
\begin{equation}
H_{jk} = \frac{ \mbox{ vol of }\mathcal{D}_j}{\mbox{ vol of the attractor of  }
\Sigma_j}, 
\end{equation}
\noindent
where vol denotes the volume \cite{explanation}. If there is no PS,
the set $\mathcal{D}_j$ spreads over the attractor of $\Sigma_j$, see
Fig. \ref{event}(a,b), then, $H_{ij}=1$. As the oscillators undergo a
transition to PS, $H_{jk}$ becomes smaller than $1$. The lower
$H_{jk}$ is the stronger the synchronization level is
\cite{commentQuio}.

\begin{figure}[!h]
  \centerline{\hbox{\psfig{file=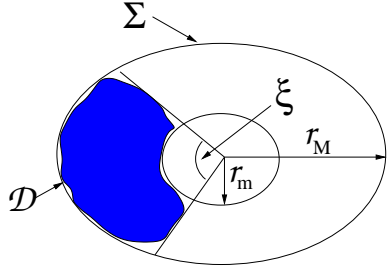,height=3.5cm}}}
\caption{ (Color online) Illustration of a localized set in a R\"ossler like attractor.
  The attractor can be approximated by a disk with major radius $r_M$
  and minor radius $r_m$. The sets $\mathcal{D}$ are confined within
  an angle $\xi$.}
\label{Hjk}
\end{figure}

For attractors with the same topology as the R\"ossler oscillator,
$H_{jk}$ can be easily calculated. Instead of computing the volume, we
calculate the area occupied by the attractor in the plane $(x,y)$. The
area $A_j$ of the attractor of $\Sigma_j$ can be roughly estimated by
the area of the disk with radii $r_m$ and $r_M$, see Fig. \ref{Hjk}.
Thus, $A_j = \pi (r_M^2 - r_m^2)$. On the other hand, the set
$\Sigma_j$ is confined into an angle $\xi$ [Fig. \ref{Hjk}].
\noindent
Therefore, the area of the set $\mathcal{D}_j$ can be estimated as
$\xi (r_M^2 - r_m^2) / 2$. Thus, the estimator can be written as:
\begin{equation}
H_{jk} = \frac{\xi}{2\pi}
\label{est_H}
\end{equation}

\begin{figure}[!h]
  \centerline{\hbox{\psfig{file=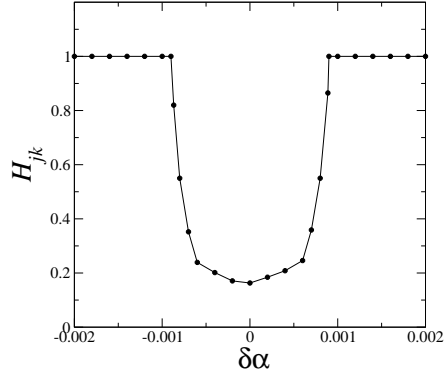,height=6cm}}}
\caption{ $H_{jk}$ is depicted  for the two coupled R\"osslers, Eq. (\ref{Rossler}),
  with $\epsilon = 0.001$. The estimator $H_{jk}$ is computed by means
  of Eq. ({\ref{est_H}}), whenever $H_{ij}=1$ there is no PS. At $|
  \delta \alpha | \approx 0.0009$ the coupled R\"osslers undergo a
  transition to PS, and therefore, $H_{jk}<1$, which shows the
  presence of PS.}
\label{Hjk_results}
\end{figure}

We have used Eq. (\ref{est_H}) to estimate the amount of
synchronization between the two coupled R\"ossler of Eq.
(\ref{Rossler}). We fix $\epsilon=0.001$ and vary the mismatch
parameter $\delta \alpha $ within the interval $[-0.002,0.002]$.  For
$| \delta \alpha | \approx 0.0009$ the coupled R\"osslers phase
synchronize, which means that the set $\mathcal{D}_j$ is localized.
Therefore, $H_{jk}<1$. The smaller the value of $| \delta \alpha |$ is
the more localized the set $\mathcal{D}_j$ becomes, meaning that the
oscillators are more synchronized, leading $H_{jk}$ to low values.  At
$| \delta \alpha | = 0$ the two coupled oscillators present their
strongest synchronization with $H_{jk} \approx 0.22$.  The results are
depicted in Fig. \ref{Hjk_results}.

\noindent

\section{Oscillators with multiple time-scales}\label{timescales}

In oscillators with only one time-scale, i.e. one typical period, a
typical event means an event possible to realize, thus with positive
measure. In oscillators with multiple time-scales, i.e. oscillators
that possess more than one typical period (an oscillator with a fast
and slow variables), a typical event means an event that takes into
account all time-scales. Conversely, an atypical event is the one that
takes into account just a few time-scales, e.g. only one.  In such an
oscillator with multiple time-scales, one may have synchronization
only in one time scale, while the others may be asynchronous.  If the
event definition excludes completely the dynamics of the synchronized
time-scale this event is atypical and one does not observe localized
sets through it.  In order to clarify these ideas, we consider two
instructive examples.

\subsection{Dynamics on a Torus}

Let us consider a quasi-periodic motion on a torus $T^2$ with two
independent frequencies $\omega$ and $\alpha$, i.e. $n \omega - m
\alpha \not= 0$ $\forall n,m\in\mathbb{Z}$. The dynamics on the torus
$\Sigma_k : T^2 \rightarrow T^2$ can be characterized by the angular
variables and the flow takes the form $\Omega_k = (u,v) = ( \omega t +
\omega_0 , \alpha t + \alpha_0 )$.  Furthermore, we consider another
oscillator on a quasi-periodic torus with two independent frequencies
$\omega$ and $\beta$, the flow $\Sigma_j: T^2 \rightarrow T^2$, in
angular variables, takes the form $\Omega_j = (g,h) = ( \omega t +
\tilde \omega_0 , \beta t + \beta_0 )$.  Therefore, under this
construction one sees PS in only one time-scale, since $\alpha,\beta$
are independent.
\begin{figure}[h]
  \centerline{\hbox{\psfig{file=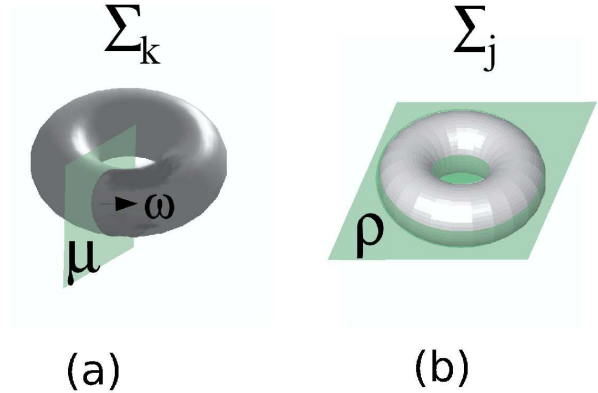,width=8.0cm}}}
\caption{ (Color online) Illustration of two possible sections on the torus $T^2$.
  In (a) the section $\mu$ takes into account only the dynamics of
  $\omega$, the synchronized scale. In (b) on the other hand, the
  dynamics of the synchronized time scale is ruled out on the section
  $\rho$. Therefore, using this particular section one cannot observe
  localized sets $\mathcal{D}$, since the synchronized time-scale is
  not taken into account.}
\label{toru}
\end{figure}
\noindent

If we consider the event in the oscillator $\Sigma_k$ to be the
increasing of $2\pi$ on the variable $u$, conversely the crossing in
the section $\mu$, it generates the times $t_k^i = 2\pi \times i /
\omega$. The observation of $\Sigma_j$ at these times generates a
localized set $\mathcal{D}_j$, which will lay on $S^1$, a subset of
$T^2 = S^1 \times S^1$, and will never occupy the full space.  On
the other hand, if we consider the event in the oscillator $\Sigma_j$
to be the increasing of $\pi$ on the variable $h$, conversely the
crossing with the section $\rho$, the set $\mathcal{D}_k$ will not be
localized, since $\alpha,\beta$ and $\omega$ are independent.

Therefore, one must define an event that captures the dynamics of the
synchronized time scale.  In the pictorial example of Fig. \ref{toru}
any other piece of section that is a linear
combination of $\mu$ and $\rho$ provides typical events.

\subsection{Spiking/Bursting Dynamics}

An interesting situation is when the time scales present a
relationship, which is the case for spiking/bursting oscillators.
Consider two spiking/bursting neurons $\mathcal{N}_j$ and
$\mathcal{N}_k$. They have distinct time-scales, the bursting scale,
with low frequencies, and the spiking scale, with high frequencies.

The spiking scale consists of the action potentials \cite{kandel}
which occur due to the exchange of ions like $K^{+}$ of the external
media with the neuron. On the other hand, the neuron may also exchange
slow current like $Ca^{+2}$ which inhibits the occurrence of spikes
generating the bursts. An event defined by the occurrence of a burst
defines simultaneously the beginning and the ending of a spike train.
Therefore, even though spikes and bursts may have independent
frequencies, the burst occurrence is also determined by the occurrence
of the first and last spike within the burst.

\begin{figure}[h]
  \centerline{\hbox{\psfig{file=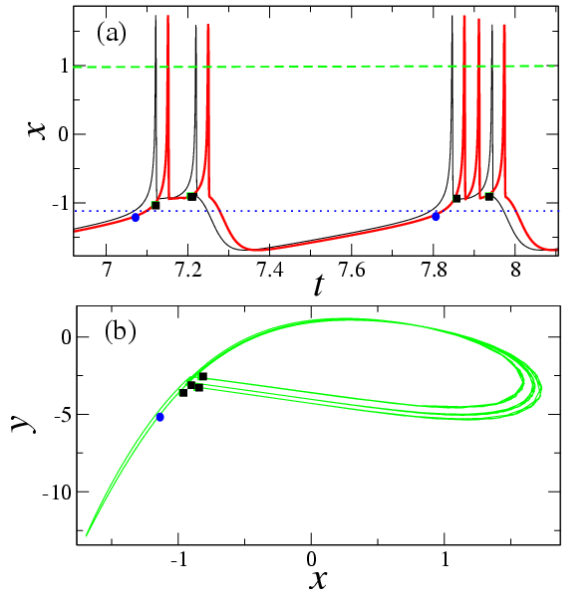,width=8.0cm}}}
\caption{ (Color online) In (a) we present the time series of the membrane
  potential of two neurons $\mathcal{N}_k$ and $\mathcal{N}_j$ in light gray
  and black respectively. We show the threshold, in dashed line, for the burst
  occurrence, and in light gray dots, for the spike occurrence. While the
  bursting scale is synchronized the spiking scale is not. However, both
  scales can be used to construct the sets $\mathcal{D}$ and they will be
  localized due to the synchronization in the bursting scale. In (b), we show
  the set $\mathcal{D}_k$ constructed using the spiking scale (black balls)
  and the set $\mathcal{D}_k$ constructed using the bursting scale (black
  squares). As one can see both are localized.}
\label{exemplo}
\end{figure}

It has been reported that it is possible to have PS in the bursting
scale while the spiking scale is not synchronized \cite{juergen}.
Therefore, in order to analyze the existence of synchronization
between the neurons, by means of standard techniques, the spiking and
bursting scales must be separately analyzed.  Our method detects PS
independently on the time-scale that the event is defined; if one
time-scale is synchronous one finds localized $\mathcal{D}$ sets.  In
order to illustrate this result we may take the following example.
Assume that the bursting scales are strongly synchronized. This means
that if neuron $\mathcal{N}_j$ ends the $i$th burst at a time $t_j^i$,
the neuron $\mathcal{N}_k$ ends the $i$th burst at a time $t_k^i =
t_j^i + \xi^i$, where $\xi^i \approx O(\eta) \ll O(1)$.  Next,
consider that within any burst in neuron $\mathcal{N}_j$ there are
always two spikes equidistant in time. Let us denote $\tau_j^n$ the
time at which the $n$th spike occurs in $\mathcal{N}_j$. In neuron
$\mathcal{N}_k$ there are two spikes within a burst and with a
probability $p_k$ a third spike may occur [ Fig.  \ref{exemplo}(a)].
Under this construction, it is clear that the spiking scales are not
synchronized.

We can verify this by applying the same approach as in Refs.
\cite{juergen,livro}.  We define a threshold for the burst occurrence,
the dot gray line in Fig. \ref{exemplo}(a). Then for every burst we
assume that the phase $\phi$ is increased by $2\pi$ and between two
bursts the phase increases linearly.  So, the phase for the neuron
$\mathcal{N}_k$ can be written as:
\begin{equation}
\phi_{k}(t) = 2\pi \times  \left( i + \frac{t - t_{k}^i}{ t_{k}^{i+1} - t_{k}^{i}} \right).
\label{pb}
\end{equation}
A similar equation can be written for $\mathcal{N}_j$.  Note only that
at a time $t$, the neuron $\mathcal{N}_j$ may present $m$ bursts.  So,
the phase difference $ | \Delta \phi| = |\phi_k(t) - \phi_j(t)|$ is
equal to $| 2\pi \times \left( i + \frac{t - t_k^i}{ t_k^{i+1} -
    t_{k}^{i}} \right) - 2\pi \times \left( m + \frac{t - t_j^{m}}{
    t_{j}^{m+1} - t_{j}^m} \right)|$. Now bringing the fact that
$|t_k^i - t_j^i| < O(\eta)$, we have
\begin{equation}
|\Delta \phi| < 4 \pi.
\end{equation}
Therefore, the phase difference is bounded. On the spiking scale the
situation is different; there is no synchronization. Doing the same
procedure, we introduce a phase $\psi$ that is increased by $2\pi$
between two successive spikes.  Thus, the phase for the neuron
$\mathcal{N}_k$ can be written as:
\begin{equation}
\psi_k(t) = 2 \pi \times \left( n + \frac{t - \tau_k^n}{ \tau_k^{n+1} - \tau_{k}}\right).
\label{spike}
\end{equation}
A naive computation in the limit $t \rightarrow \infty$ shows that
$|\Delta \psi| \approx p_k n$. Hence, there is, of course, no
synchronization on the spiking scale.

Next, we construct the set $\mathcal{D}_j$ observing the neuron
$\mathcal{N}_j$ every time that an event happens in the neuron
$\mathcal{N}_k$. First, we fix the event to be the ending of a burst,
see Fig. \ref{exemplo}(a).  As we observe $\mathcal{N}_j$ at $t_k^i$
all the points of $\mathcal{D}_j$ will be close to the end of the
burst. So, the set $\mathcal{D}_j$ does not spread over the attractor,
see the gray points in Fig. \ref{exemplo}(b).  However, the set
$\mathcal{D}_j$ is also localized even if we set the event to be the
occurrence of a spike. Since the spikes always occur within a burst,
even though the spikes themselves are not synchronized, the trajectory
related to the hyperpolarization period will not the visited, and
therefore, the set $\mathcal{D}_j$ will be localized, see the black balls in Fig. \ref{exemplo}(b).  

\noindent

\section{Neuronal Dynamics}\label{neus}

Next, we study the appearance of PS between two spiking/bursting
neurons of the Hindmarsh-Rose (HR) type. In such an oscillator the
introduction of a phase is rather difficult, since the neurons are
non-coherent.  We couple the neurons via inhibitory synapses, which
introduces non-coherence in both time-scales. This happens because
when one neuron spikes it inhibits the other neuron, which
hyperpolarizes, but the neuron that has been inhibited still tries to
spike. This competition generates even more non-coherence in both
time-scales. Therefore, we consider this model as a proper example to
illustrate our results.

In the 4-dimensional HR model \cite{reynaldo,HR} neurons are described
by a set of four coupled differential equations:
\begin{eqnarray}
\dot{x}_k &=& ay_k + bx^2_k - cx^3_k - dz_k + I_k + g_{syn} {\bf C} {\bf I}_{syn}({\bf x}) \nonumber \\
\dot{y}_k &=& e - y_k + fx^2_k - gw_k, \\
\dot{z}_k &=& \mu(-z_k + R(x_k+H)), \nonumber \\
\dot{w}_k &=& \nu(-kw_k + r(y_k+l)), \nonumber 
\label{HR_eq}
\end{eqnarray}
where $x_k$ represents the membrane potential of the neuron
$\mathcal{N}_k$, $y_k$ is associated with fast currents exchange and
$(z_k,w_k)$ with slow currents dynamics, ${\bf I}_{syn}({\bf x})=
(I_{syn}(x_1),I_{syn}(x_2),\ldots,I_{syn}(x_N))$ is the synaptic input
vector and $I_{syn}(x_j)$ is the synaptic current that neurons
$\mathcal{N}_j$ (post-synaptic) injects in $\mathcal{N}_k$
(pre-synaptic), and ${\bf C} = \{c_{kj}\}$ is the $N \times N$
connectivity matrix where $c_{kj} = 1$ if neuron $\mathcal{N}_j$ is
connected to neuron $\mathcal{N}_k$, and $c_{kj} = 0$, otherwise, with
$j \not = k$.  This model has been shown to be realistic, since it
reproduces the membrane potential of biological neurons
\cite{Johnson}, and it is able to replace a biological neuron in a
damaged biological network, restoring its natural functional activity
\cite{Mulle}. It also reproduces a series of collective behaviors
observed in a living neural network \cite{reynaldo}. The parameters of
the model are the same as in Ref.  \cite{reynaldo}, but the intrinsical
current $I_k$.  We change $I_k$ in order to obtain a spiking/bursting
behavior and we use it as a mismatch parameter.  First, we consider
two neurons $\mathcal{N}_j$ and $\mathcal{N}_k$. In the following, we
consider the parameters $I_k = 3.1200$, $I_j = 3.1205$ and
$g_{syn}=0.85$.

The chemical synapses \cite{sharp} are modeled by:
\begin{eqnarray}
I_{syn}(x_j) &=& S(t) \left( x_{rev} - x_j \right), \\
\left[ 1 - S_{\infty}(x_i) \right] \tau \dot{S}(t) &=& S_{\infty}(x_i) - S(t), \nonumber
\label{syn1}
\end{eqnarray}
where $x_j$ is the post-synaptic neuron, $x_{rev}$ is the reversal
potential for the synapse, and $\tau$ is the time-scale governing the
receptor binding.  $S_{\infty}$ is given by:
\begin{eqnarray}
S_{\infty}(x_i) = \left\{
\begin{array}{ccc}
tanh\left( \frac{x_i - x_{th}}{x_{slope}} \right), & \mbox{if} & x_i > x_{th} \\
0 & \mbox{otherwise} \\
\end{array}
\right.
\label{s2}
\end{eqnarray}
\noindent
The synapse parameters are $x_{th}=-0.80$, $x_{slope}= 1.00$,
$x_{rev}=-1.58$. They are chosen in such a way to obtain an inhibitory
effect in the chemical synapse.

\begin{figure}[h]
  \centerline{\hbox{\psfig{file=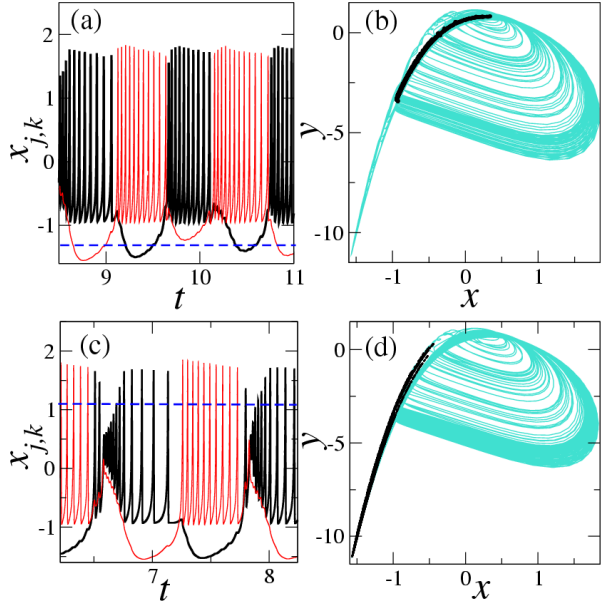,width=8.0cm}}}
\caption{ (Color Online) PS between two HR neurons coupled via inhibitory synapses. 
  We analyze the effect of different threshold levels on the detection
  of PS.  We depicted the attractor projection $(x,y)$ in gray, and
  the set $\mathcal{D}_k$ in black, for Figs. (b) and (d).  In (a) the
  time series of the membrane potentials (full lines) and the
  threshold $x_b=-1.3$ (dashed line) are depicted. In (b) the set
  $\mathcal{D}_k$, construct by means of the threshold $x_b$, is
  localized; showing the presence of PS.  The time series of the
  membrane potentials (full lines) and the threshold $x_s=1.1$ (dashed
  line) are depicted in (c). The spikes are not in PS. With our
  method, even for this threshold one can obtain a localized set.  In
  (d), the set $\mathcal{D}_k$ construct by using the threshold
  $x_s=1.1$ is localized.}
\label{ps-set}
\end{figure}
\noindent

To construct the sets $\mathcal{D}$, we define the event occurrence.
We shall analyze two situations: when the event is defined in the
bursting scale, and when the event is defined in the spiking scale.
Firstly, we define the $i$th event to be the $i$th crossing of the
membrane potential of the neuron $\mathcal{N}_{j,k}$ with the
threshold $x_b = -1.3$ in an upwards direction. We denote the time
events by $t_{j,k}^i$.  Note that this threshold assigns to the times
$t_{j,k}^i$ the beginning of the $i$th burst of $\mathcal{N}_{j,k}$.
Fig. \ref{ps-set}(a) shows the time series of the membrane potential
of the neurons $\mathcal{N}_{j,k}$. The threshold $x_b = -1.3$ is
depicted with the dashed line, and it is chosen in such a way that it
does not define a proper Poincar\'e section, which means that not all
the bursts cross it, see Fig \ref{ps-set}(a).  Actually many bursts
are missed. Thus, the approach to extract the phase considering the
increasing of $2 \pi$ between two bursts, misleads the statement of
PS.  That is so, because in this approach the phase is threshold
dependent.  Therefore, by using Eq. (\ref{pb}), we get that PS does
not exist, which is crucially wrong (note that, with the increasing of
the threshold value PS would appear).  However, our approach, which is
not threshold dependent, overcomes these difficulties.  Indeed,
localized sets exist even for this threshold [Fig \ref{ps-set}(b)].

Conversely, if we increase the threshold level in such a way that it
takes into account the spike occurrence, e.g. a threshold at $x_s =
1.1$, the dashed line in Fig.  \ref{ps-set}(c), the former approach,
as in Eq. (\ref{spike}), completely fails to state PS, due to the fact
that the spikes are not in PS. Furthermore, the spikes are highly
non-coherent.  The competition between the two neurons generates a
damping in the spikes in the beginning of the burst, followed by an
increasing and then decreasing in the spike frequency [Fig.
\ref{ps-set}(c)].  Again, since the threshold $x_s = 1.1$ defines a
typical event, the observation of $\mathcal{N}_{j,k}$ at times
$(t_{k,j}^i)_{i\in\mathbb{N}}$ provides localized sets $\mathcal{D}$ [
Fig. \ref{ps-set}(d)].

\section{Excitatory Neural Networks}\label{net}

The ideas introduced herein are also useful to analyze the onset of
synchronization in networks. We consider a network of $16$
non-identical HR neurons, regarded as $\mathcal{N}_i$ where $i\in
[1,\ldots,16]$, connected via excitatory chemical synapses.  The
mismatch parameter is the intrinsic current $I_i$. Since the
meaningful parameter is $I_i = 3.12$, for which the HR neuron best
mimics biological neurons, we introduce mismatches around this value
for all the neurons within the network. Thus, given a random number
$\eta_i$ uniformly distributed within the interval $[-0.05,0.05]$, we
set $I_i = 3.12 + \eta_i$. The excitatory synapses are modeled by Eqs.
(18) and (\ref{s2}).  To obtain the excitatory effect we
change the value of $x_{rev}$.  If $x_{rev} \ge x_i(t)$, the
pre-synaptic neuron always injects a positive current in the
post-synaptic one.  Since the maximum spike amplitude is around $1.9$,
we set $x_{rev} = 2.0$.

\begin{figure}[h]
  \centerline{\hbox{\psfig{file=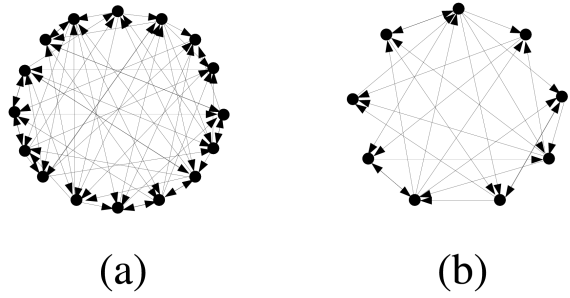,width=8.0cm}}}
\caption{ Networks generated randomly. In (a) n=16 and k=4, while  in (b) n=9 and k=3.}
\label{nets}
\end{figure}

Our network is a homogeneous random network, i.e. all neurons receive
the same number $k$ of connections, namely $k=4$, see Fig.
\ref{nets}(a).  We constrain $g_{syn}$ [see Eq. (17)] to be equal to
all neurons. We identify the amount of phase synchronous neurons by
analyzing whether the sets $\mathcal{D}_i$ are localized.
\noindent
\noindent

\begin{figure}[h]
  \centerline{\hbox{\psfig{file=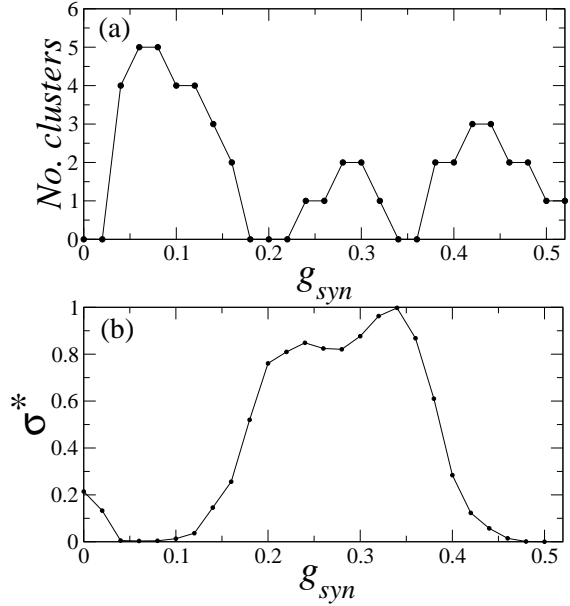,width=8.0cm}}}
\caption{ The appearance of PS clusters within the network. In (a)
  we show the number of clusters as a function of the synaptic
  strength. In (b) we plot the normalized $\sigma^*$, where $\sigma^*
  = \sigma/ 112.8$. Note that when the clusters are formed $\sigma$
  becomes small, but bounded away from zero, which means that the
  neurons within the network undergo a transition where they have
  almost the same frequency.}
\label{var_cluster}
\end{figure}

The onset of PS in the whole network takes place at $g^*_{syn} \approx
0.47$; so all neurons become phase synchronized. As the synapse
strength crosses another threshold, $\tilde g_{syn} \approx 0.525$,
the neurons undergo a transition to the rest state no longer
presenting an oscillatory behavior. Clusters of PS appear for $g_{syn}
\ll g_{syn}^*$.  In fact, right at $g_{syn} \approx0.04$, some PS
clusters appear [ Fig.  \ref{var_cluster}(a)].  Again, the clusters
are identified by analyzing the localized sets.  These clusters seem
to be robust under small perturbations.

Clusters of PS inside the network may offer a suitable environment for
information exchanging.  Each one can be regarded as a channel of
communication, since they possess different frequencies, each channel
of communication operates in different bandwidths. To see the
bandwidths in the network, we analyze the variance in the average
bursting time of the neurons. Since only the burst scale is
synchronized, we are just interested in the average bursting time,
which can be straightforwardly estimated with a fast Fourier
transformation FFT \cite{FFT}. So, given the neuron $\mathcal{N}_j$,
we label its bursting average time by $\langle T_j \rangle$. Then, we
compute the variance of the average time on the ensemble of neurons.
For this, we first introduce the average time of the whole network,
which is given by:
\begin{equation}
\zeta = \frac{1}{n}\sum_{j=1}^{n}\langle T_j \rangle.
\end{equation}
Thus, the variance of the average time on the ensemble of neurons is 
readily written as:
\begin{equation}
\sigma = \frac{1}{n} \sum_{j=1}^{n}( \langle T_j \rangle - \zeta )^2.
\end{equation}
So, $\sigma$ indicates how diverse are the bandwidths. As one can see
in Fig \ref{var_cluster}(b), when the first clusters appear for
$g_{syn}\approx 0.04$, we have $\sigma \approx 0$ indicating that the
whole network is working almost with the same frequency. A further
increasing of $g_{syn}$ causes the destruction of these clusters and
an increasing of $\sigma$.  However, even in the regimes of high
$\sigma$ with $g_{syn} \in [0.27 , 0.34]$, there is the formation of
clusters.
 \noindent
\noindent

\begin{figure}[h]
  \centerline{\hbox{\psfig{file=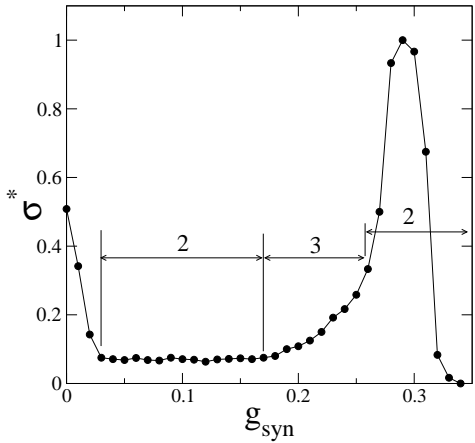,width=7.0cm}}}
\caption{ The average bursting time on the ensemble of neurons.
  We plot the normalized $\sigma^*$, where $\sigma^* = \sigma/ 12$.
  The inset numbers show the amount of clusters for a given parameter
  $g_{syn}$}
\label{hasler_PS}
\end{figure}

This scenario of cluster formation is neither restricted to this HR
model nor to the synapse model. It can also be found in square-wave
and parabolic bursters, and it is in general achieved quite before the
onset of complete synchronization. For example, we use a more
simplified HR model given by: $\dot{x}_j = a x_j^2 - x_j^3 - y_j -z_j
- g_{syn}(x_j - x_s) {\bf C}{\bf I}_{syn}({\bf x}) , \dot{y} = (a +
\alpha) x^2 - y, \dot{z} = \mu (bx +c -z)$, with the parameters:
$a=2.8, \alpha = 1.6,c=5,b=9, \mu = 0.001$; ${\bf C}$ being the
connectivity matrix and ${\bf I}_{syn}({\bf x}) = (I_{syn}(x_1) ,
\ldots, I_{syn}(x_N))$ a fast threshold modulation as synaptic input
given by
\begin{equation}
I_{syn}(x_j) = 1/[1+ exp\{ -\beta( x_j - \Theta)\}], 
\end{equation}
with $\beta=10$ and $\Theta = -0.25$. As before, $ g_{syn}$ is the
synaptic strength and the reversal potential $x_s > x_j(t)$ in order
to have an excitatory synapse. For a homogeneous random network of $9$
identical HR neurons, with $k=3$ [ Fig. \ref{nets}(b)], the theory
developed in Ref. \cite{hasler} predicts the onset of complete
synchronization at $\bar g _{syn} \approx 0.425$, while we found that
PS in the whole network is already achieved at $g^*_{syn} \approx
0.36$.  Clusters of PS, however, appear for a much smaller value of
the coupling strength, actually at $g_{syn}\approx 0.03$. Next, we
apply the same procedure as before and we compute the variance of the
average bursting time on the ensemble of neurons within the network.
The result $\sigma \times g_{syn}$ is depicted in Fig.
\ref{hasler_PS}, the inset numbers indicate the amount of clusters.

\noindent

As we have pointed out, such clusters are rather suitable for
communication exchanging mainly for two reasons: $(i)$ They have
different frequencies, therefore, each cluster may be used to transmit
information in a particular bandwidth, which may provide a
multiplexing processing of information. $(ii)$ The clusters of phase
synchronous neurons provide a multichannel communication, that is, one
can integrate a large number of neurons (chaotic oscillators) into a
single communication system, and information can arrive simultaneously
at different places of the network. This scenario may have
technological applications, e.g. in digital communication
\cite{Lai,Grebogi}, and it may also guide us towards a better
understanding of information processing in real neural networks
\cite{mormann:2003,fell:2002,realnets}.

\section{Detection of PS for higher Dimensional System}\label{High}

It is easy to say whether the set $\mathcal{D}$ is localized in a two
dimensional plane; this could be done for example by visual
inspection. In multi-dimensional system it might not be obvious
whether the set $\mathcal{D}$ is localized. This is mainly due to the
fact that in a projection of a higher dimensional system onto a low
dimensional space, the set $\mathcal{D}$ might fulfill the projected
attractor.  Therefore, the analysis of the localization might have to
be realized in the full attractor of the subsystem.

The analysis is also relatively easy if we bring about a property of
the conditional observation. Whenever there is PS, the conditional
observation, given by ${\bf F}_j^{t_k^i}({\bf x}_j^0)$, is not
topologically transitive \cite{wiggins} in the attractor of
$\Sigma_j$, i.e. $\mathcal{D}$ is localized. The conditional
observations ${\bf F}_j^{t_k^i}({\bf x}_j^0)$ are topologically
transitive in the attractor $\mathcal{A}_j$ of $\Sigma_j$
\cite{comment} if for any two open sets
$\mathcal{B},\mathcal{C}\subset \mathcal{A}$,

{\small
\begin{equation}
\exists t_k^{n_i}\text{ }/\text{ } {\bf F}_j^{t_k^{n_i}}(\mathcal{B})\cap \mathcal{C}\neq \emptyset. 
\label{transitive}
\end{equation}
\noindent} 
\noindent

To check whether $\mathcal{D}_j$ is localized, we do the following. If
there is PS, for ${\bf y}_j \in \mathcal{D}_j$ it exists infinitely
many ${\bf x}_j \in \mathcal{A}_j$ such that
\begin{equation}
{\bf y}_j \cap B_{\ell}({\bf x}_j) = \emptyset,  
\label{practice_basic_sets}
\end{equation}
\noindent
where $B_{\ell}({\bf x}_j)$ is an open ball of radius $\ell$ centered
at the point ${\bf x}_j$, and $\ell$ is small. We may vary ${\bf y}_j$
and ${\bf x}_j$ to analyze whether it is possible to fulfill Eq.
(\ref{practice_basic_sets}).  Whenever this is possible, it means that
the set $ \mathcal{D}_j$ does not spreads over the attractor of
$\Sigma_j$, and therefore, there is PS.

For analysis of PS basing on experimental data
\cite{roy,baptista:2003,Epa} where the relevant dynamical variables
can be measured, so that the phase space is recovered, our approach
can be used straightforwardly. If one just has access to a bivariate
time series, one first has to reconstruct the attractors, and then
proceed the PS detection by our approach

\section{Information Transmission}\label{info}

In this section, we analyze the relationship between the sets
$\mathcal{D}$ and the capacity of information transmission between
chaotic oscillators.  In order to proceed such an analysis, we may
assume that the oscillators are identical or nearly identical. Such
that the synchronized trajectories are not far from the
synchronization manifold, i.e. the subspace where ${\bf x}_j = {\bf
  x}_k$ \cite{pecora90,pecora98}.  Next, for a sake of simplicity we
consider only oscillators whose trajectory possess a proper rotation
and are coherent \cite{Lai,Josic}, e.g.  the standard R\"ossler
oscillator. However, the ideas herein can be extended to other
oscillators as well.

The amount of information that two systems $\Sigma_j$ and $\Sigma_k$
can exchange is given by the mutual information $I(\Sigma_j,\Sigma_k)$
\cite{Shannon}:
\begin{equation}
I(\Sigma_j,\Sigma_k) = H(\Sigma_j) -  H(\Sigma_j|\Sigma_k),
\end{equation}
\noindent
where $H(\Sigma_j)$ is the entropy of the oscillator $\Sigma_j$ and
$H(\Sigma_j|\Sigma_k)$ is the conditional entropy between $\Sigma_j$
and $\Sigma_k$ , which measures the ambiguity of the received signal,
roughly speaking the errors in the transmission.

As pointed out in Ref. \cite{Murilo-Canal} the mutual information can
be also estimated through the conditional exponents associated to the
synchronization manifold. The mutual information is given by:
\begin{equation}
I(\Sigma_j,\Sigma_k) = \sum \lambda^{+}_{\parallel} - \sum \lambda^{+}_{\perp} 
\end{equation}
where $\lambda^{+}_{\parallel}$ are the positive conditional Lyapunov
exponents associated to the synchronization manifold, the information
produced by the synchronous trajectories, and $\lambda^{+}_{\perp}$
are the positive conditional Lyapunov exponents transversal to the
synchronization manifold, related with the errors in the information
transmission.  In PS $\lambda^{+}_{\perp}$ can be small, which means
that one can exchange information with a low probability of errors.
So, PS creates a channel for reliable information exchanging
\cite{Murilo-Canal}.  In general, we expect $\sum
\lambda^{+}_{\parallel} \le \sum \lambda^{+}$, where $\lambda^{+}$ are
the positive Lyapunov exponents.  Thus $I(\Sigma_j,\Sigma_k) \le \sum
\lambda^{+} - \sum \lambda^{+}_{\perp}$.  In order to estimate an
upper bound for $I(\Sigma_j,\Sigma_k)$, we need to estimate
$\lambda^{+}_{\perp}$, what can be done directly from the localized
sets.

The conditional transversal exponent can be estimated from the
localized sets by a simple geometric analysis. At the time $t_j^i$ the
oscillator $\Sigma_j$ reaches the Poincar\'e plane at ${\bf x}_j^*$
while the oscillator $\Sigma_k$ is at ${\bf x}_k^i = {\bf x}_k(t_j^i
)$. The initial distance between the trajectories is $\Delta {\bf
  x}_{jk} = {\bf x}_j^* - {\bf x}_k^i$. This distance evolves until
the time $t_k^i$ when the oscillator $\Sigma_k$ reaches the Poincar\'e
plane at $x_k^*$, while the trajectory of $\Sigma_j$ is at ${\bf
  x}_j^i = {\bf x}_j(t_k^i )$.  The new distance is $\Delta \tilde
{{\bf x}}_{jk}(t_k^i-t_j^i) ={\bf x}_k^* - {\bf x}_j(t_k^i)$.
Therefore, we have:

\begin{equation}
\Delta \tilde{{\bf x}}_{jk} = \Delta {\bf x}_{jk} e^{\lambda_{\perp}^{+} |t_k^i - t_j^i|}
\end{equation}

So, the local transversal exponent is given by:
\begin{equation}
\lambda_{\perp}^{+} = \lim_{N \rightarrow \infty}\frac{1}{N} \sum_{i=1}^{N}\frac{1}{|t_j^i - t_k^i|} 
\ell n  \left| \frac{{\bf x}_j^* - {\bf x}_k^i}{{\bf x}_k^* - {\bf x}_j^i}\right|,
\label{lambda}
\end{equation}
\noindent
where we use the convention $0 \times log 0=0$.  Of course, we only
estimate the conditional exponent close to the Poincar\'e plane.
Hence, if we change the Poincar\'e plane the conditional exponent may
also change, i.e. there are some events that carry more information
than others.
 
\subsection{Example with R\"ossler Oscillators}

We illustrate this approach for two coupled R\"ossler oscillators. We
set the parameters to $a=0.15$, $b=0.2$, $c=10$, $\alpha_j =1$, and
$\Delta \alpha_k = 0.0002$. As shown in Ref.  \cite{Murilo-Canal} at
$\epsilon \approx 0.05$, the two oscillator undergo a transition to
PS. In particular, for $\epsilon = 0.06$ we have $\sum
\lambda_{\perp}^{+} \approx 0.06$. We estimate $\sum
\lambda_{\perp}^{+}$ at this situation by means of Eq.
(\ref{lambda}).  We set the Poincar\'e section at $y_{j,k}=0$, and
compute $\lambda_{\perp}$ for $65,000$ cycles, i.e. 65,000 crossing of
the trajectory with $y=0$ and $\dot{y}<0$.  We get $\lambda_{\perp}
\approx 0.048$. Note that we are not computing $\sum
\lambda_{\perp}^{+}$, but rather, the maximum $\lambda_{\perp}^{+}$,
namely $\tilde{\lambda}_{\perp}^{+} $. Therefore, it is natural to
expect $\lambda_{\perp}^{+}$ to be smaller than $\sum
\lambda_{\perp}^{+}$.  However, the upper bound to the information
exchange can be estimated by $I(S,R) \le \sum \lambda^{+} - \tilde
{\lambda}^{+}_{\perp}$, that is, the maximum amount of information
that can flow through the coupled oscillators if we encode the
trajectory using the Poincar\'e plane $y=0$ \cite{Grebogi}.
Furthermore, it seems that when the level of synchronization is large,
the estimation of $\tilde {\lambda}^{+}_{\perp}$, by means of Eq.
(\ref{lambda}), might become problematic, due to strong fluctuations
in $|t_j^i - t_k^i|^{-1} \ell n [|({\bf x}_j^* - {\bf x}_k^i)/({\bf
  x}_k^* - {\bf x}_j^i)|]$.

\section{Conclusions}\label{conc}

We have proposed an extension of the stroboscopic map, as a general
way to detect PS in coupled oscillators.  The idea consists in
constraining the observation of the trajectory of an oscillator at
these times in which typical events occur in the other oscillator.
This approach provides an efficient and easy way of detecting PS,
without having to explicitly calculate the phase.  We have shown that
if PS is present, the maps of the attractor appear as a localized set
in the phase-space.  This has been illustrated in
coherent oscillators, the coupled R\"osslers, as well as in
non-coherent oscillators, spiking/bursting neurons of HR type coupled
with chemical synapses. As we have shown in neural networks, the
appearance of clusters of PS is rather common, which may be relevant
for communication mainly due to two aspects: $(i)$ The clusters
provide multiplexing information processing, namely each cluster may
be used to transmit information within a bandwidth.  $(ii)$ They
provide a multichannel communication, that is, a large number of
neurons is integrated into a single communication system.  Moreover,
we have analyzed the relation between the information exchanging and
the localized sets. We have roughly estimated the errors in the
information transmission from the localized sets.

{\bf Acknowledgment} We would like to thank M. Thiel, M. Romano, C.
Zhou, and L. Pecora for useful discussions. This work was financially
supported by the Helmholtz Center for Mind and Brain Dynamics, EU COST
B27 and DFG SPP 1114.

\appendix

\section{Proof of Theorem 1}\label{proof}

In this appendix we prove the theorem 1. It is instructive to give a
sketch of the proof, in order to have a better understanding of the
result.  We split the demonstration into the following four steps:
$(i)$ We show that the increasing of $2 \pi$ in the phase $\phi_{j,k}$
defines a smooth section $\Gamma_{j,k}$ on $\Sigma_{j,k}$, which does
not intersect itself.  $(ii)$ We show that observing the oscillator
$\Sigma_j$ whenever oscillators $\Sigma_k$ crosses $\Gamma_k$ gives
place to a localized set $\mathcal{D}_j$. $(iii)$ Further, we show
that the observation of $\Sigma_j$ whenever $\Sigma_k$ crosses a piece
$P_{\Gamma_k}$ of the section $\Gamma_k$ also gives place to a
localized set. $(iv)$ Using these results we show that, actually, the
localized sets can be constructed using any typical event.  To show
this, we only note that given a typical event with positive measure,
we can choose $P_{\Gamma_k}$ to be close to the event occurrence,
implying that shortly before or shortly after of every event
occurrence, a crossing of the trajectory with $P_{\Gamma_k}$ will
happen. Thus, if we observe $\Sigma_j$ whenever the event occurs in
$\Sigma_k$ we will have a set that a close the $\mathcal{D}_j$, and
therefore, localized.  Next, we formalize the heuristic ideas. Let us
introduce $i=j,k$.

\begin{propo}
  The increasing of $2 \pi$ in $\phi_{i}(t)$ generates a smooth
  section $\Gamma_{i}$ in the attractor of $\Sigma_{i}$, which
  does not intersect itself.
\end{propo}

{\it Proof:} Firstly, let us introduce the times $(\tau_{i}^m)$ such
that $\phi_{i}(\tau_{i}^m) = m \times 2 \pi$.  Then, let
$\Gamma_{i}$ be the set of points such that given the initial point
${\bf x}_{i}^{\ell}$ we have the section:
\begin{equation}
\Gamma_{i} = \{ \cup_{m \in\mathbb{N}} {\bf x}_{i}^m \, | \, {\bf
  x}_{i}^m = {\bf F}_{i}^{\tau_{i}^m} ({\bf x}_{i}^{\ell})\}
\end{equation}

Thus, we construct a section $\Gamma_{i}$. $\Gamma_{i}$ is smooth
since both $\phi_{i}$ and ${\bf F}_{i}^t$ are smooth. Indeed,
given two points ${\bf x}_{i}^0,{\bf x}_{i}^1 \in \Gamma_{i}$,
with $d( {\bf x}_{i}^0 , {\bf x}_{i}^1) < \epsilon$, there is a $r
\ge 1$ such that ${\bf F}_{i}^{\tau_{i}^r}({\bf x}_{i}^0), {\bf
  F}_{i}^{\tau_{i}^r }({\bf x}_{i}^1) \in \Gamma_{i}$, and
\begin{equation}
d({\bf F}_{i}^{\tau_{i}^r}({\bf x}_{i}^0), {\bf
  F}_{i}^{\tau_{i}^r}({\bf x}_{i}^1)) < \delta.
\end{equation}
\noindent

Furthermore, we can construct a continuous section $\Gamma_{i}$, by
conveniently choosing points ${\bf x}_{i}^{\ell}$.  The fact that
$\Gamma_{i}$ does not intersect itself comes from the uniqueness of
${\bf F}_{i}^t$ \cite{teoremaunicidade}, and from the fact that the
$\dot{\phi}_{i}(t) > 0$, which implies that the phase is an one-to-one
function with the trajectory. Note that, obviously, this section
depends on the initial conditions. $\Box$

\begin{lemma}
  The observation of the oscillators $\Sigma_j$ whenever the
  trajectory of $\Sigma_k$ crosses the section $\Gamma_k$ gives place
  to a localized set ${\mathcal{D}_j}$ if, and only if, there is PS.
\end{lemma}

{\it Proof:} Let $\Pi_j$ be the Poincar\'e map associated to the
section $\Gamma_j$, such that given a point ${\bf x}_j^{n} \in
\Gamma_j$, so ${\bf x}_j^{n+1} = \Pi_j({\bf x}_j^{n})$ = $F_j^{\Delta
  \tau_j^{n+1}}({\bf x}_j^n)$, where $\Delta \tau_j^{n}$ = $\tau_j^{n}
- \tau_j^{n-1}$.  From now on, we use a rescaled time $t^{\prime} = t
/ \langle T_j \rangle $, with $\langle T_j \rangle =
\lim_{i\rightarrow \infty} \tau_j^i/i$.  For a slight abuse of
notation we omit the $\prime$. There are  numbers $\kappa_{i}$ such
that $| \tau _{i} ^{i} - i \langle T_{i} \rangle | \leq
\kappa_{i},$ where, by time reparametrization, $\kappa_{i} \ll 1$.
If both oscillators are in PS, then $\langle T_k \rangle = \langle T_j
\rangle$, and so:
\begin{equation}
| \tau_k^{n} - \tau_j^{n}| \leq \tilde \kappa,
\label{diferenca_temporal}
\end{equation}
\noindent 
with $\tilde \kappa \le \kappa_k + \kappa_j \ll 1$ \cite{commentN}.
Now, we analyze one typical oscillation, using the basic concept of
recurrence. Given the following starting points ${\bf x}_k^0 \in
\Gamma_k$ and ${\bf x}_j^0 \in \Gamma_j$, we evolve both until ${\bf
  x}_j^0$ returns to $\Gamma_j$.  Let us introduce
\begin{equation}
\Delta \tau^n = \Delta \tau_j^n - \Delta \tau_k^n.
\end{equation}

Which gives:
\begin{equation}
{\bf F}_j^{\Delta \tau_j^{1}}({\bf x}_j^0) = \Pi_j({\bf x}_j^0)=
{\bf x}_j^1 \in \Gamma_j.
\end{equation}
Analogously, 

\begin{eqnarray}
{\bf F}_k^{\Delta \tau _j^{1}}({\bf x}_k^0) &=& {\bf F}_k^{ \Delta \tau_k^{1} + \Delta \tau^1}({\bf x}_k^0) \nonumber\\
&=& {\bf F}_k^{ \Delta \tau^1} \circ {\bf F}_k^{ \Delta \tau_k^{1}}({\bf x}_k^0). \nonumber
\end{eqnarray}
Bringing the fact that ${\bf F}_k^{ \Delta \tau_k^{1}}({\bf x}_k^0) =  \Pi_k({\bf x}_k^0) = {\bf x}_k^1$, we have:
\begin{equation}
{\bf F}_k^{\Delta \tau _j^{1}}({\bf x}_k^0) = {\bf F}_k^{ \Delta  \tau^1}({\bf x}_k^1).
\end{equation}
Now, by using the fact that $| \Delta \tau^i | < \tilde \kappa $, we
can write:
\noindent
{\small
\begin{equation}
 {\bf F}_k^{ \Delta \tau^1}({\bf x}_k^1) \approx {\bf x}_k^1 + {\bf
 G}({\bf x}_k^1) \tilde \kappa + \mathcal{O}(\tilde \kappa ^2).
 \label{tiagoIII}
\end{equation}}
\noindent
So, given a point ${\bf x}_k \in \Gamma_k$ evaluated by the time when
the trajectory of $\Sigma_j$ returns to the section $\Gamma_j$, the
point ${\bf x}_k$ returns near the section $\Gamma_k$, and vice-versa.
Therefore, it is localized. For a general case, we have to show that a
point, in the section $\Gamma_k$, evolved by the flow for an arbitrary
number $N$ of events in the oscillator $\Sigma_j$, still remains close
to $\Gamma_k$, in other words, it is still localized.  This is
straightforward, since $|\sum_{i=0}^N \Delta \tau^i| = |\tau_k^N -
\tau_j^N | < \tilde \kappa$. So, we demonstrated that the PS regime
implies the localization of the set $\mathcal{D}_k$.

Now, we show that the localization of the set $\mathcal{D}_k$ implies
PS. Supposing that we have a localized set $\mathcal{D}_k$, so, Eq.
(\ref{diferenca_temporal}) is valid, by the above arguments.
Therefore, we just have to show that Eq.  (\ref{diferenca_temporal})
implies PS.  With effect, we have $| \phi_j(t) - \phi_k(t) | = |
\int_0^t \Omega_j dt - \int_0^t \Omega_k dt |$ which is equal to $ |
\int_0^{\tau_j^n} \Omega_j dt - \int_0^{\tau_j^n} \Omega_k dt +
\int_{\tau_j^n}^t \Omega_j dt - \int_{\tau_j^n}^t \Omega_k dt |$. This may
be written as $| \int_0^{\tau_j^n} \Omega_j dt - \int_0^{\tau_k^n} \Omega_k
dt - \int_{\tau_j^n}^{\tau_k^n} \Omega_k dt + \int_{\tau_j^n}^t \Omega_j dt -
\int_{\tau_j^n}^t \Omega_k dt |$. Next, noting that
$\phi_{i}(\tau_{i}^n) = 2\pi \times n$, we get:
\begin{equation}
| \phi_j(t) - \phi_k(t) | \le  M | \tau_j^n - \tau_k^n | + 2 \Lambda M, \\
\end{equation}
where $ \Lambda = max|t_{i}^n - t_{i}^{n-1}|$.  Therefore, if the time
event difference $| t_j^n - t_k^n |$ is bounded it implies the
boundedness in the phase.  Thus, we conclude our result.  $\Box$

\begin{propo}
  Let $( \tau_j^{n_i})_{{n_i}\in\mathbb{N}}$ be the times at which the
  trajectory $\Sigma_j$ crosses a piece $P_{\Gamma_j}$ of $\Gamma_j$.
  If there is PS, then the observation of the trajectory of $\Sigma_k$
  at times $(\tau_j^{n_i})_{n_i\in\mathbb{N}}$ gives place of a
  localized set.
\end{propo}

{\it Proof:} Note that the observation of the trajectory of $\Sigma_k$
at times $(\tau_j^i)_{i\in\mathbb{N}}$ gives place to a set
$\mathcal{D}_k$, while the observations at times {\small
  $(\tau_j^{n_i})_{n_i\in\mathbb{N}}$} give place to a subset
$\tilde{\mathcal{D}}_k$ of $\mathcal{D}_k$. Therefore, whenever
$\mathcal{D}_k$ is localized, it implies the localized of
$\tilde{\mathcal{D}}_k$. $\Box$

 \noindent
\begin{figure}[h]
  \centerline{\hbox{\psfig{file=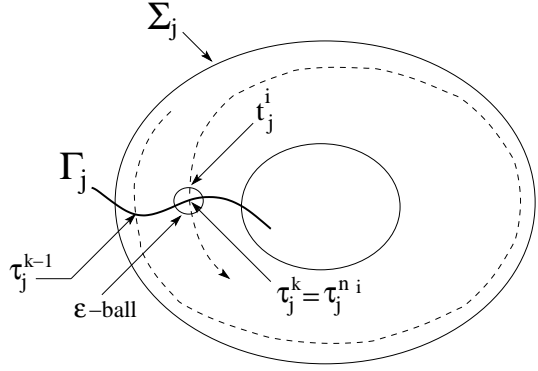,width=7.0cm}}}
\caption{ Illustration of Eq. (\ref{prova}).}
\label{example}
\end{figure}
\noindent
    
Now, we are ready to prove the $theorem$ 1.

{\it Proof:} Let the event be the entrance in an $\varepsilon$-ball,
such that the event occurrence produces the time series $t_j^i$, in
$\Sigma_j$.  There is, at least, one intersection of this ball with
the section $\Gamma_j$.  Since $\Gamma_j$ depends on the initial
conditions, we can choose an initial condition right at the
$\varepsilon$-ball event.  Next, we choose $P_{\Gamma_j}$ such that it
is completely covered by the $\varepsilon$-ball. Since the measure of
the $\varepsilon$-ball is small, $\varepsilon \ll 1$, the time
difference between crossings of the trajectory with $P_{\Gamma_j}$ and
the $\varepsilon$-ball is small, thus, there is a number $\eta < 1$
such that:
\noindent
\begin{equation}
t_j^i - \tau_j^{n_i} \ll O(\eta).
\label{prova}
\end{equation} 
\noindent
Therefore, if we observe the trajectory of $\Sigma_k$ at times
$(t_j^i)_{i\in \mathbb{N}}$, we have a localized set in $\Sigma_k$.
Thus, we conclude our result: The observation of the trajectory of
$\Sigma_{j,k}$ whenever typical events in $\Sigma_{k,j}$ occurs
generates localized sets $\mathcal{D}_{j,k}$ if, and only if, there is
PS.  $\Box$


\begin{thebibliography}{99}
\bibitem{explanation} We consider vol to be the euclidian volume that
  contains all the points of the set $\mathcal{D}$. Thus, if the set
  $\mathcal{D}$ is spread over a sphere of radius $r$ the volume of
  the set $\mathcal{D}$ will be $\pi r^3$. The same is valid for the
  attractor volume.
\bibitem{commentQuio} Note the the choice of $H_{jk}$ is somewhat
  arbitrary. We could also compare the euclidian distance between the
  points of the attractor and the points of the localized set. This
  would lead to the approach of Ref. \cite{quio}.
\bibitem{comment} A chaotic set is always transitive through the flow.
  So, given a set of initial conditions, its evolution through the
  flow eventually reaches arbitrary open subsets of the original
  chaotic attractor.  However, the conditional observations might not
  possesses the transitive property.
\bibitem{commentN} We could also show that $\tilde \kappa$ is smaller
  than 1 by the following argument. Note that the number of crossings
  $N_j$ and $N_k$ of the oscillators $\Sigma_j$ and $\Sigma_k$ with
  the sections $\Gamma_j$ and $\Sigma_k$, respectively, may not be
  always the same.  They can differ by an unity, since the events are
  not simultaneous. Therefore, we have $| N_j(t) - N_k(t)| \leq 1$.
  However, $N_i \approx \lim_{t\rightarrow \infty} t / \langle T_i
  \rangle$, then, using this into the previous equation and
  normalizing $\langle T_j \rangle = 1$, we have $| \tau_j^{n} -
  \tau_k^{n}| \le O(1)$, for $n$ large enough.
\end{thebibliography}
\end{document}